\documentclass[conference]{IEEEtran}
\IEEEoverridecommandlockouts

\usepackage[T1]{fontenc}
\usepackage[utf8]{inputenc}
\usepackage{float}
\usepackage{amsmath,amssymb}
\usepackage{graphicx}
\graphicspath{{figures/}}
\usepackage[short]{optidef}
\usepackage{xcolor}
\usepackage{hyperref}
\hypersetup{colorlinks,urlcolor=blue,citecolor=.,linkcolor=.}
\usepackage{siunitx}
\usepackage[thinc]{esdiff}
\usepackage{multirow}

\usepackage{silence}
\usepackage{subcaption}

\renewcommand{\t}{^\top}
\renewcommand{\matrix}[1]{\begin{bmatrix}#1\end{bmatrix}}
\newcommand{\T}[1]{\Tilde{#1}}

\newcommand{\m}[1]{\ensuremath{\mathrm{#1}}}

\usepackage{tikz}
\usetikzlibrary{patterns}
\newcommand{\poleangle}{30}

\usepackage{amsthm}
\newtheorem{definition}{Definition}

\begin{document}

\title{Direct transfer of optimized controllers to similar systems using dimensionless MPC}

\author{Josip Kir Hromatko$^1$, Shambhuraj Sawant$^2$, Šandor Ileš$^1$, Sébastien Gros$^2$

\thanks{
$^1$Josip Kir Hromatko and Šandor Ileš are with the Faculty of Electrical Engineering and Computing, University of Zagreb, Croatia.\newline E-mail: \texttt{\{josip.kir.hromatko, sandor.iles\}@fer.unizg.hr}

$^2$Shambhuraj Sawant and Sébastien Gros are with the Department of Engineering Cybernetics, Norwegian University of Science and Technology (NTNU), Trondheim, Norway.\newline E-mail: \texttt{\{shambhuraj.sawant, sebastien.gros\}@ntnu.no}

$^*$This work was supported in part by the Croatian Science Foundation.}
}

\maketitle

\begin{abstract}
    Scaled model experiments are commonly used in various engineering fields to reduce experimentation costs and overcome constraints associated with full-scale systems. The relevance of such experiments relies on dimensional analysis and the principle of dynamic similarity. However, transferring controllers to full-scale systems often requires additional tuning. In this paper, we propose a method to enable a direct controller transfer using dimensionless model predictive control, tuned automatically for closed-loop performance. With this reformulation, the closed-loop behavior of an optimized controller transfers directly to a new, dynamically similar system. Additionally, the dimensionless formulation allows for the use of data from systems of different scales during parameter optimization. We demonstrate the method on a cartpole swing-up and a car racing problem, applying either reinforcement learning or Bayesian optimization for tuning the controller parameters. Software used to obtain the results in this paper is publicly available at \url{https://github.com/josipkh/dimensionless-mpcrl}
\end{abstract}


\section{Introduction}
Since the early days of control theory, there has been a natural progression toward tackling increasingly complex problems.
As new control methods mature, interest in both research and practice tends to shift toward, for example, larger systems or systems with greater uncertainty.
However, in some cases, experimenting on the real system might be prohibitively dangerous or expensive.

Scaled model experimentation is a well-established engineering practice across many fields \cite{coutinho2015}.
An underlying assumption is that the full-size system and the scaled system have a certain degree of similarity.
This notion is formalized using dimensional analysis and the Buckingham $\Pi$ theorem \cite{buckingham}, which defines the conditions under which two systems are \textit{dynamically similar}.
When dynamic similarity is ensured, experiments done on a lab-scale system provide results analogous to full-scale testing, which can complement numerical simulations (especially with systems that are difficult to model).

While scaled models offer additional experimental flexibility, finding the optimal controller for a given task might not be straightforward, especially in more advanced control methods such as model predictive control (MPC).
To reduce the effect of model inaccuracies or inherent system uncertainty, there is growing interest in combining classical control methods with machine learning tools \cite{ml_control_survey,piml_survey,mesbah2022}.
One promising framework that combines the strengths of both fields is the integration of MPC with reinforcement learning (RL) \cite{rlmpc_survey_2025}.
The performance-driven nature of RL and the strong theoretical foundation of MPC provide an appealing combination of safety and closed-loop optimality, as demonstrated in recent literature \cite{gros_data_driven,gros2022,romero2024,brandner2024}.
A parallel approach, which has also received attention recently, is to automatically adjust controller parameters for closed-loop performance using Bayesian optimization (BO) \cite{piga2019,sorourifar2021}.

However, directly transferring controllers optimized for performance between different systems, referred to as \textit{zero-shot transfer}, can be challenging and lacks theoretical guarantees \cite{kostas21}.
Without a formal notion of system similarity, there is often a compromise between generalization capabilities and closed-loop performance.
Consequently, most practical approaches rely on online fine-tuning to recover acceptable performance \cite{valassakis2020crossing}.
Dimensional analysis provides a structured framework for studying dimensionally similar systems and enables the design of a single controller that is guaranteed to be optimal for an entire range of (in theory, infinitely many) similar systems.

In \cite{brennan2001} and \cite{brennan2005}, the authors present a robust controller design method based on dimensional analysis, applied to the vehicle steering problem.
The goal is to develop a controller in a dimensionless space and enable its immediate transfer to dynamically similar vehicles of different sizes.
Similarly, dynamic matching was used in \cite{solc2022} to accurately scale a full-size vehicle to laboratory dimensions.
Despite its potential benefits, dimensional analysis appears to have received limited attention in control system design.
In parallel, leveraging dimensional analysis for controller transfer has recently seen renewed interest in the machine learning community.
It was used for system identification in \cite{bakarji2022}, while \cite{pascoa2025} discusses transferring existing policies between dynamically similar systems.
Improving the generalization capability of a reinforcement learning policy using \textit{dimensionless Markov decision processes} (MDPs) has recently been suggested in \cite{charvet2025}.
This relates to \textit{contextual MDPs} \cite{hallak2015}, where key system parameters are incorporated alongside observations for determining the optimal policy.
However, a seamless transfer of the policy and closed-loop performance between systems of different scales is ensured only if the associated MDPs satisfy certain similarity requirements.

In this paper, we first extend the concept of dimensionless MDPs and introduce the notion of \textit{similar MDPs}, formalizing the conditions for a direct policy transfer.
We then present a method of nondimensionalizing a nonlinear MPC scheme, which becomes invariant to the scale of the underlying system.
Finally, the dimensionless MPC scheme is used to approximately solve a (theoretically infinite) range of similar MDPs through their dimensionless formulation.
We illustrate the applicability of our approach with two examples, a cartpole swing-up and a car racing task, using either RL or BO to tune the MPC parameters for closed-loop performance.

\section{Background}
\subsection{Markov decision processes}
MDPs are a modelling framework for discrete-time stochastic decision processes \cite{puterman_mdp}. 
An MDP is commonly defined by a tuple:
\begin{equation}\label{eq:mdp}
    \mathcal{M} \coloneqq \big(\mathcal{S}, \mathcal{A}, \mathcal{P}, \mathcal{L},\gamma\big)
\end{equation}
which contains a state space $\mathcal{S} \subseteq \mathbb{R}^{n_s}$ defining the possible states $s$, an action space $\mathcal{A} \subseteq \mathbb{R}^{n_a}$ defining the possible actions $a$, a transition kernel given by the conditional probability density $\mathcal{P}(s^+|s,a)$, a stage cost of each state-action pair $\mathcal{L}(s,a)$ and a discount factor $\gamma$, representing the relative reduction in costs incurred further in the future.

An MDP is approximately solved by finding an optimal policy $\pi_\theta^*(s)$, parametrized by $\theta$ for practical tractability, which minimizes the expected cost $J$ across all states:
\begin{equation}
    \pi_\theta^* = \arg \min_\theta J(\pi_\theta)
\end{equation}
with the objective defined as:
\begin{equation}
    J(\pi_\theta)=\mathbb{E}_{s\sim\rho^{\pi_\theta}}\left[\left. \sum_{t=0}^\infty\gamma^t\mathcal{L}(s_t,a_t) \right| a_t\sim\pi_\theta(\cdot | s_t) \right]
\end{equation}
and $\rho^{\pi_\theta}$ denoting the state distribution under policy $\pi_\theta$.

\subsection{Solving MDPs using model predictive control}
Finding the optimal policy for an MDP can be done in several ways.
In \cite{gros_data_driven}, the authors show that an MPC scheme is a valid function approximator and can be used as a policy candidate. A parametrized, discounted nonlinear MPC scheme can be expressed as:
\begin{mini!}
    {x,u}{\gamma^NT_\theta(x_N) + \sum_{k=0}^{N-1}\gamma^kL_\theta(x_k,u_k)\label{eq:mpc_cost}}{\label{eq:param_mpc}}{}
    \addConstraint{
    x_0=s_t \label{eq:mpc_x0}}{}{}
    \addConstraint{
    x_{k+1}=f_{\m{d},\theta}(x_k,u_k) \label{eq:mpc_model}}{}{\quad k=0 ,\ldots ,N -1}
    \addConstraint{
    h_\theta(x_k,u_k)\leq 0\label{eq:mpc_constraint}}{}{\quad k=0 ,\ldots ,N -1}
    \addConstraint{
    h_{N,\theta}(x_k)\leq 0\label{eq:mpc_constraint_e}}{}{}
\end{mini!}
where $x=\{x_0,x_1,\dots,x_{N}\}$, $u=\{u_0,u_1,\dots,u_{N-1}\}$, $N$ denotes the number of steps in the prediction horizon, $s_t$ is the current state of the system, $L_\theta$ and $T_\theta$ represent the stage and terminal cost, $f_{\m{d},\theta}$ denotes the discrete-time model of the system dynamics and $h_\theta$, $h_{N,\theta}$ contain the stage and terminal constraints.
State and input predictions are denoted as $x_k$ and $u_k$, while $\gamma$ represents the discount factor.
The solution of the problem gives the optimal control sequence $u^*$, and its first element becomes the MPC policy, $\pi_\theta(s_t)=u_0^*$.

Compared to the more common neural networks, using an MPC scheme as a function approximator allows for a more direct embedding of physical models and, in our case, system similarities.
In this paper, we will focus on using MPC as the underlying policy $\pi_\theta$ and RL or BO for policy optimization.
Both methods offer tradeoffs between global optimality and scalability with the number of parameters $\theta$ \cite{perez2022}.

\subsection{Dimensional analysis}
\label{sec:dimensional_analysis}

Similarities between systems of different scales are often used by engineers to simplify models, reduce the cost of experimentation, and facilitate the analysis of system characteristics.
Scale model experiments, often performed in fields such as aerospace and naval engineering, rely on dimensional analysis to ensure an appropriate match between the scaled and the full-sized systems.

A central theorem in dimensional analysis was formulated by E. Buckingham \cite{buckingham}, and it states that any equation of the form:
\begin{equation}
\label{eq:buckingham_dimensional}
    g(q_1,q_2,q_3,\dots,q_n)=0
\end{equation}
can be rearranged into the form:
\begin{equation}
\label{eq:buckingham_dimensionless}
    \T{g}(\Pi_1,\Pi_2,\Pi_3,\dots,\Pi_{n-r})=0
\end{equation}
where $\Pi_i$ are scalar dimensionless terms formed from scalar dimensional quantities $q_i$.
In practice, this transformation (often called \textit{nondimensionalization}) is typically carried out as follows:
\begin{enumerate}
    \item List all dimensional quantities $q_i$ and their corresponding physical units (e.g., in the SI unit system).
    \item Identify the $r$ fundamental dimensions involved (e.g., mass, length, and time in mechanical systems).
    \item From the set of $q_i$, select $r$ dimensionally independent repeating variables that collectively include all the fundamental dimensions (e.g., a characteristic mass, length, and time constant).
    \item Form $n-r$ dimensionless groups $\Pi_i$ by multiplying the remaining dimensional variables with appropriate powers of the repeating variables, $\Pi_i=q_i\cdot\prod_{j=n-r+1}^n q_j^{\alpha_j}$.
    \item Express the original equation in terms of the dimensionless groups $\Pi_i$.
\end{enumerate}

Once they are formed, the dimensionless terms $\Pi_i$ can provide insight into connections between variables $q_i$ and their effect on the system dynamics.
It is also possible for two sets of dimensional quantities $\{q_i\}_{i=1}^n$ to lead to the same dimensionless terms, $\{\Pi_i\}_{i=1}^{n-r}$.
In that case, the systems defined by the two dimensional sets are described by the same dimensionless equation and are said to be \textit{dynamically similar}.
Consequently, given a set of reference dimensional parameters, one can find the corresponding parameters of a dynamically similar system of different scale by matching the dimensionless terms.
For more details and applications of dimensional analysis, we refer the reader to \cite{balaguer2013}.
\section{Dimensionless and similar MDPs}
As presented in \cite{charvet2025}, the idea of nondimensionalization can be extended to MDPs. In this section, we elaborate the procedure of obtaining \textit{dimensionless MDPs} and introduce the notion of \textit{similar MDPs}.

\subsection{Dimensionless MDPs}
The transition kernel of an MDP, $\mathcal{P}(s^+|s,a)$, can be expressed in a functional form with an explicit dependency on a stochastic disturbance $d$ and fixed system parameters $p$:
\begin{equation}
\label{eq:mdp_trans}
    s^+ = F(s, a, d; p)
\end{equation}
which can be written in the form \eqref{eq:buckingham_dimensional} with $q=\{s^+,s,a,d,p\}$. By inspecting the physical units of $s$, $a$ and $d$, we can select a subset of $p$ as the repeating variables for dimensional analysis, as described in Section \ref{sec:dimensional_analysis}.
Then, the process of nondimensionalization boils down to a linear transformation of the variables:
\begin{equation}
\label{eq:mdp_scaling}
    s=M_s\T{s},\quad a=M_a\T{a},\quad d=M_d\T{d}
\end{equation}
where $\T{s}$, $\T{a}$, $\T{d}$ are dimensionless variables and $M_s$, $M_a$, $M_d$ are diagonal, positive definite matrices containing only groups of the repeating variables. Each group in these matrices forms the scaling factor for the corresponding dimensional variable. Using \eqref{eq:mdp_scaling} in \eqref{eq:mdp_trans} and rearranging the expressions yields a dimensionless transition model:
\begin{subequations}
\label{eq:mdp_trans_nondim}
\begin{align}
    \T{s}^+ &= M_s^{-1}\cdot F(M_s\T{s},M_a\T{a},M_d\T{d};p) \\
    &=  \T{F}(\T{s}, \T{a}, \T{d}; \T{p}) 
\end{align}    
\end{subequations}
which can be written in the form \eqref{eq:buckingham_dimensionless} with $\Pi=\{\T{s}^+,\T{s},\T{a},\T{d},\T{p}\}$.
Note that the dimensional parameters $p$ still enter the expressions \eqref{eq:mdp_trans_nondim}, but will now be lumped into dimensionless terms $\T{p}$ (otherwise the expressions would not be dimensionally consistent, see the examples in \cite{balaguer2013}).
For details on applying transformations to random variables, please see \cite[Section 3.6.2]{rice2007}.
In the particular case of a normally distributed disturbance, $d\sim\mathcal{N}(\mu_d,\Sigma_d)$, the transformed disturbance will be distributed as $\T{d}\sim\mathcal{N}(M_d^{-1}\mu_d, M_d^{-1}\Sigma_d(M_d^{-1})^\top$. 

Dimensionless states $\T{s}$ and actions $\T{a}$ belong to the dimensionless state and action spaces:
\begin{subequations}
\label{eq:mdp_spaces_scaling}
\begin{align}
    \T{\mathcal{S}} &= \{\T{s}\in\mathbb{R}^{n_s} | \T{s}=M_s^{-1}s, s\in\mathcal{S}\} \\
    \T{\mathcal{A}} &= \{\T{a}\in\mathbb{R}^{n_a} | \T{a}=M_a^{-1}a, a\in\mathcal{A}\}
\end{align}
\end{subequations}
while the stage cost $\mathcal{L}$ can be transformed using \eqref{eq:mdp_scaling}:
\begin{equation}
\label{eq:mdp_cost_scaling}
    \T{\mathcal{L}}(\T{s},\T{a})=m_\mathcal{L}^{-1}\cdot\mathcal{L}(M_s\T{s},M_a\T{a})
\end{equation}
where the scalar $m_\mathcal{L}$ is formed from the repeating variables and indicates the physical unit of the reward, similar to the matrices in \eqref{eq:mdp_scaling}. The discount factor $\gamma$ is already dimensionless, so it does not change in the MDP reformulation. Finally, a dimensionless version of a dimensional MDP \eqref{eq:mdp} can be defined as a tuple:
\begin{equation}\label{eq:mdp_nondim}
    \T{\mathcal{M}} \coloneq \big(\T{\mathcal{S}}, \T{\mathcal{A}}, \T{\mathcal{P}}, \T{\mathcal{L}},\gamma\big)
\end{equation}
where $\T{\mathcal{P}}(\T{s}^+|\T{s},\T{a})$ is the probabilistic form of \eqref{eq:mdp_trans_nondim}.

\subsection{Similar MDPs}
Analogously to the dynamically similar systems in Section \ref{sec:dimensional_analysis}, we can formalize the notion of similarity between MDPs:
\begin{definition}[Similar MDPs]
    The dimensional MDPs defined in the form \eqref{eq:mdp} are said to be similar MDPs if they share the same dimensionless form \eqref{eq:mdp_nondim}.
\end{definition}
If a pair of MDPs is similar, solving one of them gives the optimal policy for the other one (with the appropriate state and action transformations).
Similar to dynamic matching in scale model design, when given a reference MDP in a dimensional form, it is possible to define a similar dimensional MDP by matching its dimensionless form.
Additionally, when considering MDPs related to similar continuous-time systems of different scales, achieving exact similarity might require a modification of the sampling time used to obtain the state transitions.

\section{Dimensionless MPC}
In this section, we present the process of nondimensionalizing an MPC problem \eqref{eq:param_mpc}, where we omit the subscript $\theta$ for simplicity. We show that in such a formulation, the problem (and its solution) are invariant to the system scale and can be used as an optimal policy candidate for the dimensionless MDP \eqref{eq:mdp_nondim}.

\subsection{Dimensionless prediction model}
\label{sec:mpc_model_nondim}
Continuous-time systems under consideration in engineering applications are often described by deterministic models in the form of ordinary differential equations (ODEs):
\begin{equation}
\label{eq:ode_dim}
    \diff{x}{t}=f(x,u;p)
\end{equation}
containing physical quantities as states $x$, inputs $u$, time $t$ and parameters $p$. Similar to \eqref{eq:mdp_scaling}, we can introduce a linear transformation of the variables:
\begin{equation}
\label{eq:scaling}
    x=M_x\Tilde{x},\quad u=M_u\Tilde{u},\quad t=m_t\T{t}
\end{equation}
where $M_x$, $M_u$, and $m_t$ are again formed using dimensional analysis and the chosen repeating variables.

Using \eqref{eq:scaling} in \eqref{eq:ode_dim} and rearranging the terms leads to the system dynamics expressed in a dimensionless form:
\begin{equation}\label{eq:ode_nondim}
    \diff{\Tilde{x}}{\Tilde{t}} = \Tilde{f}(\Tilde{x}, \Tilde{u}; \T{p})
\end{equation}
with the new, dimensionless variables as states, inputs, and time. If two systems are dynamically similar (i.e., have the same $\Pi$-groups), their physical models \eqref{eq:ode_dim} will be different, but the dimensionless ones \eqref{eq:ode_nondim} identical. For more details and examples of nondimensionalizing ODEs, please see \cite{Conejo2021}. Discretizing \eqref{eq:ode_nondim} with the dimensionless sampling time:
\begin{equation}
\label{eq:dt_nondim}
    \Delta\T{t}=m_t^{-1}\Delta t
\end{equation}
with $\Delta t$ denoting the chosen sampling time of the controller, gives a dimensionless discrete-time model of the system dynamics:
\begin{equation}
\label{eq:de_nondim}
    \T{x}_{k+1}=\T{f}_\m{d}(\T{x}_k,\T{u}_k;\T{p})
\end{equation}
which can be used to obtain the state predictions in the MPC scheme.

\subsection{Dimensionless cost and constraints}
The MPC cost and constraints can be expressed in relative terms by applying \eqref{eq:scaling} to the original cost \eqref{eq:mpc_cost}:
\begin{subequations}
\label{eq:mpc_cost_nondim}
\begin{align}
    \tilde{L}(\tilde{x},\tilde{u}) &= m_L^{-1}\cdot L(M_x\tilde{x}, M_u\tilde{u}) \\
    \tilde{T}(\tilde{x}) &= m_T^{-1}\cdot T(M_x\tilde{x})
\end{align}
\end{subequations}
and constraints \eqref{eq:mpc_constraint}:
\begin{equation}
\label{eq:h_nondim}
    \T{h}(\T{x},\T{u})=M_h^{-1}\cdot h(M_x\tilde{x}, M_u\tilde{u})
\end{equation}
The additional terms $m_L$, $m_T$ are scalars and $M_h$ is a diagonal matrix defining the physical units of the cost and constraints, similar to those used for transformations in \eqref{eq:scaling} and \eqref{eq:mdp_cost_scaling}. Note that $m_L=m_T=1$ in typical tracking costs, since the nondimensionalization is implied in the cost weights\footnote{For example, an MPC cost $L(x,u)=q_1\cdot(x_1-x_\m{1,ref})^2+q_2\cdot(x_2-x_\m{2,ref})^2$ with unspecified units of $q_i$ is only meaningful if the units of $x_1$ and $x_2$ are compatible. In the general case, dimensional consistency is ensured by defining $q_i=\T{q}_i/m_i^2$ where $m_i$ carries the physical units of $x_i$ and $\T{q}_i$ is dimensionless. With this reformulation, $L(x,u)$ will be dimensionless, regardless of the units of $x$ and $u$.}. Finally, the dimensionless MPC formulation requires a dimensionless current state estimate, obtained from the physical state estimate as:
\begin{equation}
    \T{s}_t=M_x^{-1}s_t
\end{equation}
The optimal dimensional control input can then be obtained as:
\begin{equation}
    u_0^*=M_u\tilde{u}_0^*
\end{equation}
where $\tilde{u}_0^*$ denotes the optimal dimensionless control input, i.e., the first element in the solution of the dimensionless MPC problem.

As with similar MDPs, a pair of dimensional MPC problems with matching dimensionless forms shares the same (dimensionless) optimal solution. Consequently, finding satisfactory controller parameters for one system (either manually or through a policy optimization method such as RL or BO) yields the corresponding parameters for the second system, with a direct transfer of closed-loop performance.

\subsection{Importance of dynamic similarity}
Exact dynamic similarity between systems of different scales can be difficult to achieve in practice, due to unmodeled dynamics or scale-dependent effects.
However, the effect of the parametric mismatch depends on the specific dimensionless group and can be estimated using sensitivity analysis.
Dimensional analysis also offers a measure of similarity between systems (e.g., distance in the $\Pi$-group space), which can be used to estimate the applicability of an existing controller to a new system. 
Finally, the properties of classic model predictive control naturally extend to the dimensionless formulation, where a minor parametric uncertainty might have a negligible effect on the closed-loop performance.

\section{Experiments}
In this section, we present the results obtained by automatically tuning the parameters of a dimensionless MPC controller using a policy optimization method.
The systems under consideration are deterministic and nonlinear, while the MPC controllers have partially parametrized cost functions.
Although the full MPC parametrization \eqref{eq:param_mpc} offers additional flexibility in parametric constraints and prediction model, automatically tuning the cost parameters is one of the more common use cases in practice.
In both examples, the same dynamics model was used for predictions in the MPC and for the simulation.
Also, we set $\gamma=1$ in both examples due to their episodic nature.
Since some details of the experiments have been omitted here for the sake of brevity, we refer to the publicly available code (link provided in the abstract) for additional information.

\subsection{Pendulum on a cart}
\subsubsection{Problem description}
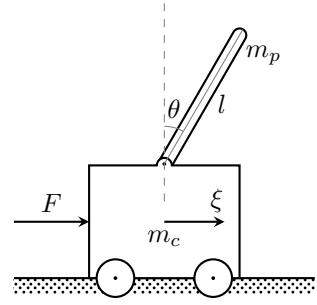
\begin{figure}
    \centering
    \begin{tikzpicture} [thick]
    \draw (-2,0) -- (2,0);
    \fill [pattern = crosshatch dots,
        pattern color = black] (-2,0) rectangle (2,-.2);
    \begin{scope} [draw = black,
        fill = white!20, 
        dot/.style = {black, radius = .025}]
        \filldraw [rotate around = {-\poleangle:(0,1.5)}] (.09,1.5) -- 
            node [midway, right] {$l$} 
            node [very near end, right] {$m_p$}
            +(0,2) arc (0:180:.09) 
            coordinate [pos = .5] (T) -- (-.09,1.5);
        \filldraw (-1,1.5) -- coordinate [pos = .5] (F)
            (-1,.0) -- node [above = .3cm] {$m_c$}
            (1,.0) -- (1,1.5) 
            coordinate (X) -- (.1,1.5)
            arc (0:180:.1) -- (-1.014,1.5);

        \filldraw (-.65,.0) circle (.25);
        \fill [dot] (-.65,0) circle;
        \filldraw (.65,0) circle (.25);
        \fill [dot] (.65,0) circle;
        \fill [dot] (0,1.52) circle;
    \end{scope}
    \begin{scope} [thin, white!50!black]
        \draw (T) -- (0,1.52) coordinate (P);
        \draw [dashed] (P) + (0,-.5) -- +(0,2.2);
        \draw (P) + (0,.5) arc (90:90-30:.5) node [black, midway, above] {$\theta$};
    \end{scope}
    \draw [stealth-] (F) -- node [above] {$F$} + (-1,0);

    \coordinate (A) at (0,0.75);
    \draw [-stealth] (A) -- node [very near end, above] {$\xi$} + (.8,.0);
\end{tikzpicture}
    \caption{The cartpole system from \cite{anand2023}.}
    \label{fig:cartpole}
\end{figure}

As a first example, we consider a modified version of the cartpole system presented in \cite[Section IV.B]{anand2023}.
It represents a pendulum mounted on top of a cart moving in one dimension, as shown in \figurename~\ref{fig:cartpole}.
The goal is to swing the pendulum up and keep it upright by controlling only the force that affects the lateral motion of the cart. 

We define the four states as $\xi$ (cart position), $\phi$ (pole angle) and their derivatives, $s=[\xi, \phi, \dot{\xi}, \dot{\phi}]^\top$.
The control input (i.e., action) is the force, $a=F$, moving the cart sideways.
Also, there are five parameters: cart and pole masses $m_\m{c}$ and $m_\m{p}$, pole length $l$, friction coefficient $\mu_\m{f}$ and the gravitational constant $g$.
For a more detailed description of the system, please see \cite{anand2023}.

\subsubsection{Dynamic matching}
The fundamental dimensions appearing in the equations are mass, length and time (which is common in mechanical systems).
Using $m_\m{c}$, $l$ and $g$ as the repeating variables gives the dimensionless groups formed from the remaining parameters:
\begin{equation}\label{eq:cartpole_pi}
    \Pi_1=\frac{m_\m{p}}{m_\m{c}},\quad \Pi_2=\frac{\mu_\m{f}}{m_\m{c}}\sqrt{\frac{l}{g}}
\end{equation}
In order to identify dynamically similar systems, we use a set of reference values for the five parameters and assume that $\mu_f$ and $g$ cannot be changed (which is often the case in practice).
Then, a system dynamically similar to the reference one can be found by selecting a new pole length $l$ and using the equivalence of the dimensionless terms \eqref{eq:cartpole_pi} for the two systems to obtain new parameters $m_\m{p}$ and $m_\m{c}$.

\subsubsection{Dimensionless MDP}
The transformation matrices in \eqref{eq:mdp_scaling} and \eqref{eq:dt_nondim} can be formed using the selected repeating variables:
\begin{equation}
\label{eq:cartpole_mdp_scaling}
    M_s = \textrm{diag}(l,1,\sqrt{gl},\sqrt{g/l}),\ \ M_a=m_\m{c}g,\ \ m_t=\sqrt{l/g}
\end{equation}
where $d$ has been omitted since the system is deterministic.
A dimensionless transition model can be obtained by applying \eqref{eq:cartpole_mdp_scaling} and the procedure described in Section \ref{sec:mpc_model_nondim} to the original cartpole model in the form of an ODE.
In our example, the MDP stage cost is defined as:
\begin{equation}
    \mathcal{L}(s,a)=-\frac{|\pi-|\phi||}{10\pi}
\end{equation}
which is dimensionless, leading to $m_\mathcal{L}=1$.
The sampling time and the state and action spaces are scaled using \eqref{eq:cartpole_mdp_scaling} in \eqref{eq:dt_nondim} and \eqref{eq:mdp_spaces_scaling}.

\subsubsection{Dimensionless MPC}
In the MPC formulation, we use the same dimensionless discrete-time model as in the MDP, with $x=s$ and $u=a$.
The original constraints on the cart position and input force can be expressed in the form \eqref{eq:mpc_constraint}:
\begin{equation}
\label{eq:h_cartpole}
    h(x,u)=\matrix{\xi-\xi_\m{max} \\ \xi_\m{min}-\xi \\ F-F_\m{max} \\ F_\m{min}-F} \leq 0
\end{equation}
The physical dimensions of the constraints suggest that $M_h$ in \eqref{eq:h_nondim} should be defined as:
\begin{equation}
    M_h=\textrm{diag}\left(l,l,m_\m{c}g,m_\m{c}g\right)
\end{equation}
When \eqref{eq:h_nondim} is applied to \eqref{eq:h_cartpole}, the dimensionless constraints become:
\begin{equation}
    \T{h}(\T{x},\T{u})=\matrix{\T{\xi}-\frac{\xi_\m{max}}{l} \\ \frac{\xi_\m{min}}{l}-\T{\xi} \\ \T{F}-\frac{F_\m{max}}{m_\m{c}g} \\ \frac{F_\m{min}}{m_\m{c}g}-\T{F}} \leq 0
\end{equation}
A similar procedure can be applied to normalize the terminal constraints on the cart position, $h_N(x)$.
Additionally, the original MPC cost:
\begin{subequations}
\begin{align}
    T(x) &= (x_\m{ref}-x)\t Q(x_\m{ref}-x) \\
    L(x,u) &= T(x) + u\t R u
\end{align}
\end{subequations}
is transformed by applying \eqref{eq:mpc_cost_nondim}:
\begin{subequations}
\label{eq:cartpole_mpc_cost_scaling}
\begin{align}
    \T{T}(\T{x}) &= (\T{x}_\m{ref}-\T{x})\t \T{Q}(\T{x}_\m{ref}-\T{x}) \\
    \T{L}(\T{x},\T{u}) &= \T{T}(\T{x}) + \T{u}\t \T{R} \T{u}
\end{align}
\end{subequations}
with $M_x=M_s$, $M_u=M_a$, $\T{Q}=M_x\t QM_x$ and $\T{R}=M_u\t RM_u$.
Note that the original cost does not carry units, which implies that $m_L=m_T=1$. The sampling time is again scaled using \eqref{eq:dt_nondim}.

\subsubsection{Experiment setup and results}
For the experiment setup, we use the cartpole example from the open-source library \texttt{Learning for predictive control (leap-c)} \cite{leap-c}, which is built on top of \texttt{CasADi} \cite{casadi} and \texttt{acados} \cite{acados}.
The MPC controller in the example has a relatively short horizon ($N=5$), which is not sufficient to successfully swing the pendulum up.
In order to circumvent this issue, a hierarchical control structure (see \figurename 3 in \cite{rlmpc_survey_2025}) is used.
In this formulation, a neural network is prepended to the MPC controller and sets the pole angle reference ($\phi_\m{ref}$) based on the current state of the system, providing additional flexibility through a state-dependent parametrization.
The reference for the remaining states is set to zero and a bounded random perturbation is added to the output of the network, in order to encourage exploration.
Since the only nonzero reference is an angle (already dimensionless), no transformation of $x_\m{ref}$ is required.
To adjust the neural network weights, we apply the soft actor-critic (SAC) method \cite{sac}.
A more detailed exposition of the method is currently under preparation.

\begin{figure}
    \centering
    \includegraphics[width=\columnwidth]{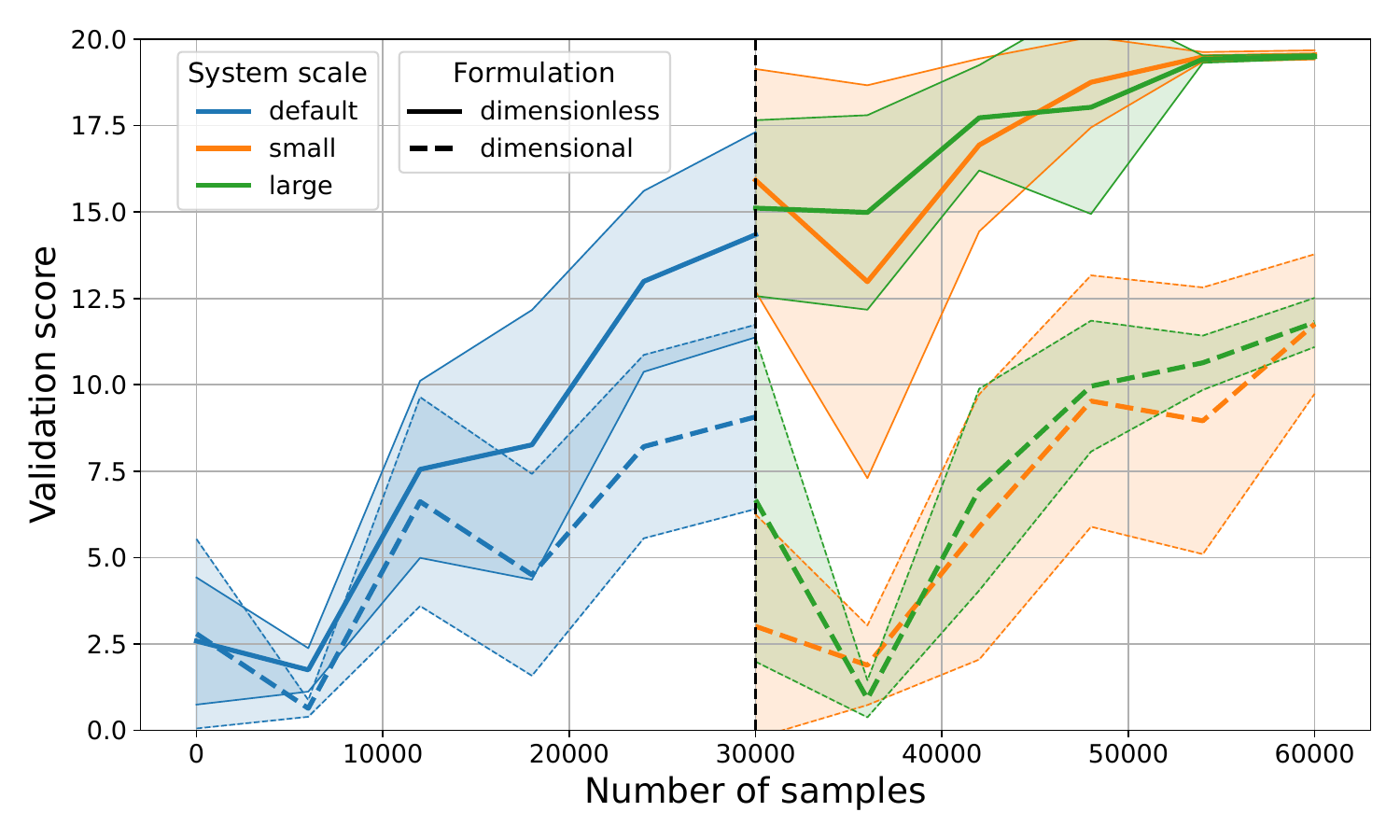}
    \caption{The progress of validation score, showing the mean and standard deviation of 5 independent runs. Note that the score corresponds to the negative MDP cost (i.e., higher is better). A value above ca. 18 indicates a successful swing-up and balancing.}
    \label{fig:cartpole_learning}
\end{figure}

To test the robustness of parameter optimization against changes in the MDP, we change the scale of the system after a fixed number of samples.
We switch from the original system ($l=0.8$ m) to either a smaller ($l=0.1$ m) or a larger ($l=5.0$ m) dynamically similar one.
The results obtained with the dimensional and the dimensionless controller formulation are shown in \figurename\ \ref{fig:cartpole_learning}.
The dimensionless formulation enables a continuation of score improvement using data from different systems, as well as a direct transfer of the optimal parameters to new, similar ones (not shown in the figure).
On the other hand, the performance of the dimensional controller is practically reset after the change.
Even when trained to convergence (not shown here), it fails to perform the swing-up as soon as the scale of the system is changed slightly.
Note that the difference in performance directly after the switch, exhibited with both controllers, should be attributed to the stochasticity of the policy.

\subsection{Race car}
\subsubsection{Problem description}
As a second example, we consider the car racing problem presented in \cite{kloeser2020}, with the goal of minimizing lap time on a known track.
In order to facilitate real-time optimization, the authors reformulate the time minimization problem as a progress maximization problem.
In this paper, we consider the case without additional obstacles on the track.
Also, we start the simulation from a standstill.

The car's pose is projected on the track centerline and described by four states: progress along the centerline, here denoted as $\sigma$, lateral distance to the centerline $n$, heading error w.r.t. the road curvature $\alpha$ and the car's speed $v$.
Control inputs are the duty cycle of the motor's PWM signal $D$ and the steering angle $\delta$.
Additionally, there are 8 parameters describing the chassis ($m$,$l$,$l_\m{r}$) and traction force generation ($c_\m{m1}$,$c_\m{m2}$,$c_\m{r0}$,$c_\m{r2}$,$c_\m{r3}$).
Please see \cite{kloeser2020} for a full description of the problem.

\subsubsection{Dynamic matching}
Through inspection of the variables appearing in the model, we can see that the fundamental dimensions are again mass, length and time.
For the repeating variables, we choose the mass $m$ and length of the car $l$, with the additional parameter $c_\m{r3}$ (chosen arbitrarily).
By using dimensional analysis, we can obtain the $\Pi$-groups:
\begin{equation}
\begin{aligned}
    & \Pi_1 = \frac{l_r}{l},\ && \Pi_2 = \frac{c_\m{m1}c_\m{r3}^2l}{m},\ &&& \Pi_3 = \frac{c_\m{m2}c_\m{r3}l}{m} \\
    & \Pi_4 = \frac{c_\m{r0}c_\m{r3}^2l}{m},\ && \Pi_5 = \frac{c_\m{r2}l}{m} &&&
\end{aligned}
\end{equation}
which can be used to identify a dynamically similar vehicle, based on the chosen (new) repeating variables.

\subsubsection{Dimensionless MDP}
The transformation matrices in this example become:
\begin{equation}
\label{eq:racecar_scaling}
    M_s = \textrm{diag}\left(l,l,1,\frac{1}{c_\m{r3}}\right),\ \ M_a=I_2,\ \ m_t=lc_\m{r3}
\end{equation}
with $I_2$ denoting an identity matrix of size 2. 
The dimensionless transition model is obtained in the same manner as in the cartpole example.
Since the aim is to minimize the total lap time, the MDP stage cost can be defined as:
\begin{equation}
    \mathcal{L}(s,a)=\Delta t
\end{equation}
leading to $m_\mathcal{L}=m_t$.
The sampling time and the state and action spaces are again normalized using \eqref{eq:racecar_scaling} in \eqref{eq:dt_nondim} and \eqref{eq:mdp_spaces_scaling}.

\subsubsection{Dimensionless MPC}
For the MPC formulation, which is defined using control input rates $\Delta u$ instead of the absolute values $u$ (see \cite{kloeser2020} for a detailed explanation), the transformation matrices are slightly modified:
\begin{equation}
    M_x^{\Delta u} = \textrm{diag}\left(l,l,1,\frac{1}{c_\m{r3}},1,1\right),\ \ M_u^{\Delta u}=\frac{1}{lc_\m{r3}}\cdot I_2
\end{equation}
The (tracking) MPC cost is scaled as in \eqref{eq:cartpole_mpc_cost_scaling}, while normalizing the constraints can be done similar as in the previous example and is omitted here for the sake of brevity.

\subsubsection{Experiment setup and results}

To illustrate the dimensionless approach with a different policy optimization method, the cost weights on the first three states ($\sigma$, $n$ and $\alpha$) were optimized using Gaussian process-based Bayesian optimization through the open-source framework Optuna \cite{optuna}.
Unlike the RL approach in the previous example, BO does not require MPC sensitivity information, which can make it a more appealing method for controller tuning.
In this example, the diagonal stage and terminal cost matrices were parametrized independently.
Additionally, we assume no access to data from the full-sized system during the optimization.

\begin{figure}
    \centering
    \includegraphics[width=.85\linewidth]{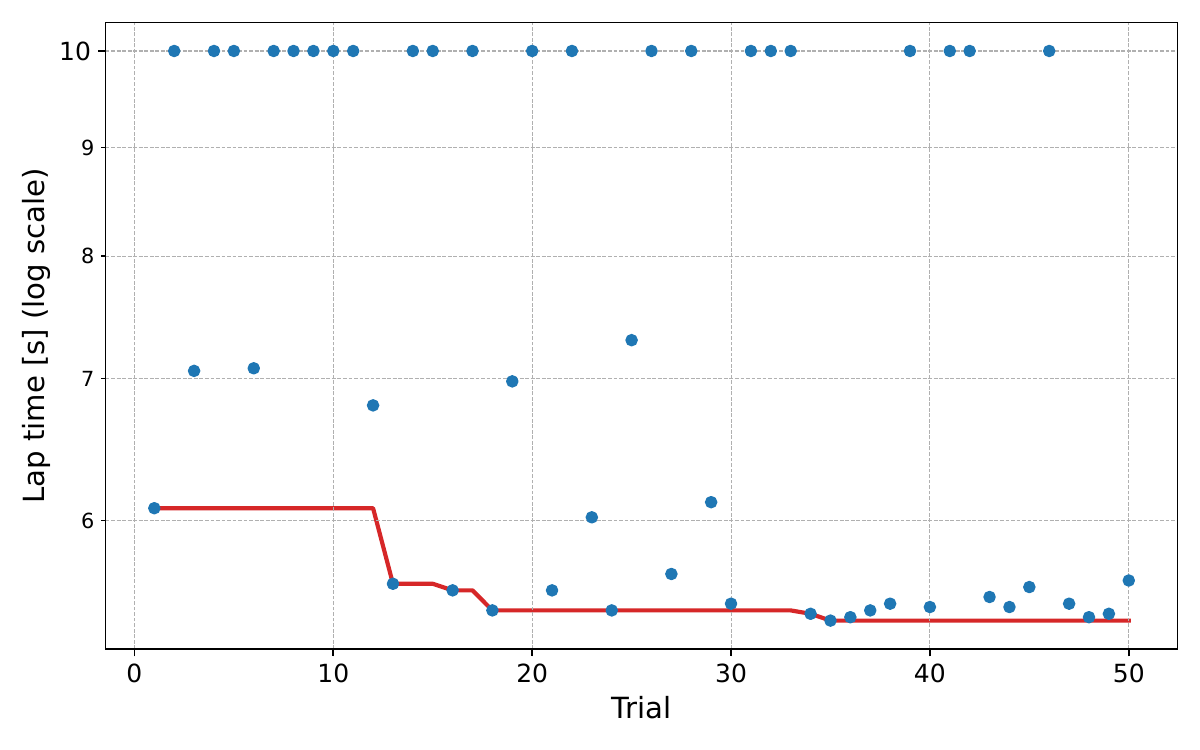}
    \caption{Results of 50 trials using the small-scale vehicle, with a time limit of 10 seconds. The red line indicates the best lap time obtained up to the specific trial.}
    \label{fig:bo_small}
\end{figure}

\begin{figure}
    \centering
    \includegraphics[width=\linewidth]{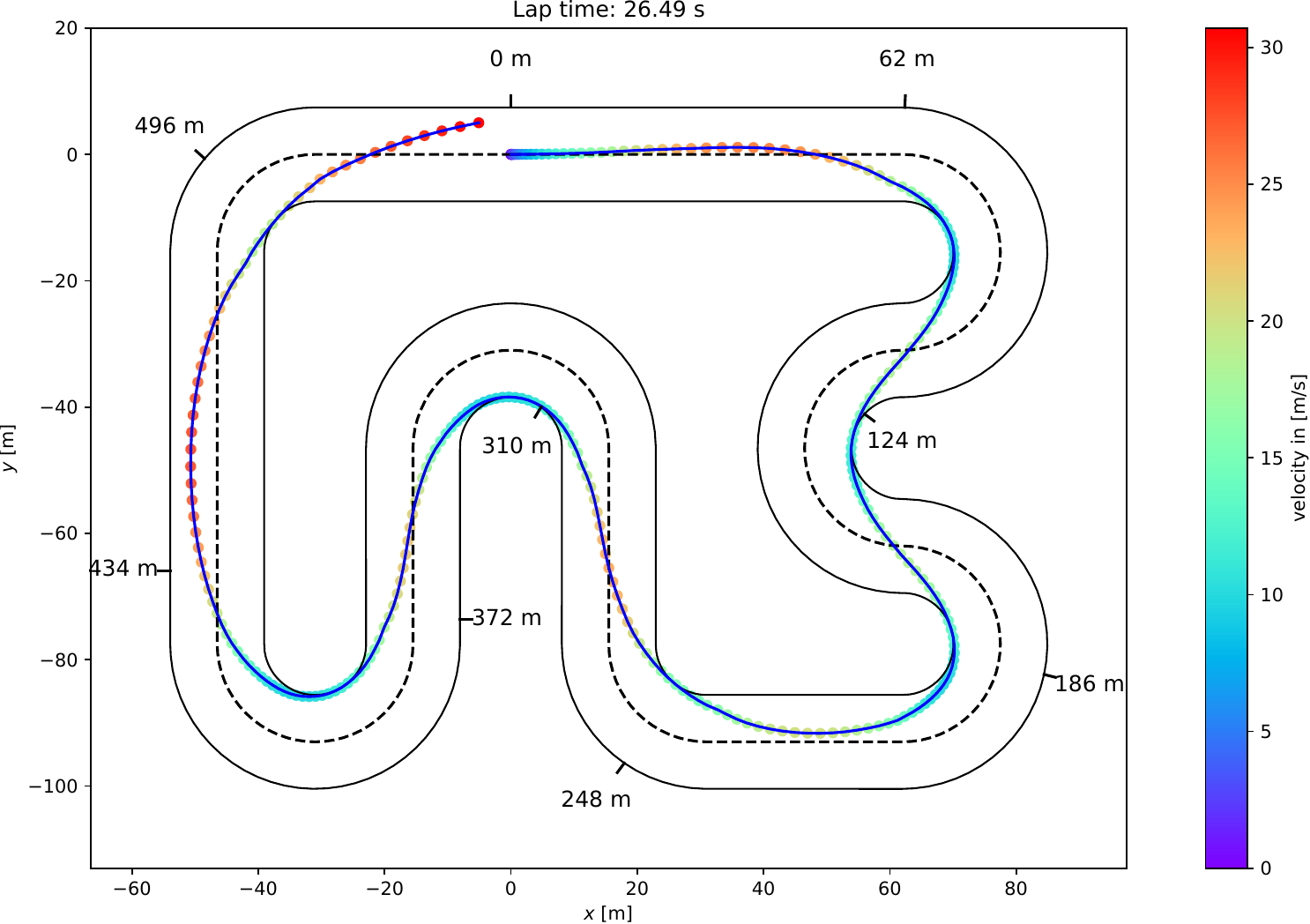}
    \caption{Full-size vehicle trajectory using the parameters optimized on a small-scale vehicle.}
    \label{fig:car_track}
\end{figure}

\figurename~\ref{fig:bo_small} shows the parameter optimization progress using the small vehicle, with $l=0.06$ m and $m=0.043$ kg.
Using the optimized parameters with a larger vehicle ($l=4$ m, $m=1500$ kg) and a scaled race track immediately results in the trajectory shown in \figurename~\ref{fig:car_track}.
Note the similarity with \figurename~1 in \cite{kloeser2020}, showing the time-optimal race line.

\section{Conclusion}
The presented approach to solving similar MDPs using an MPC-based policy in a dimensionless formulation effectively condenses a given problem, leading to a solution which is optimal for an entire class of similar problems (with the appropriate coordinate transformations).
This formulation enables a more direct transfer of controllers between systems of different scales, while retaining the closed-loop properties.
Also, the approach enables continuous and seamless learning from data generated by different systems, which can reduce the total number of full-scale experiments required to complete a given task.

\section*{Acknowledgement}
J.K.H. would like to thank Filippo Airaldi and Valentin Charvet for helpful discussions on the contents of this paper.

\bibliographystyle{IEEEtran}
\bibliography{content/bibliography}

\end{document}